\documentclass[10pt]{Procedia_PIUTAM}

\usepackage{times}
\usepackage{mathptmx}
\global\autotrue
\usepackage{subfigure}

\usepackage{graphicx}
\graphicspath{{figs/}{figs_lo/}}
\usepackage[fleqn]{amsmath}
\usepackage{amsfonts}
\usepackage[square,numbers]{natbib}
\usepackage{fleqn}
\usepackage[letter]{crop}

\usepackage{hyperref}

\usepackage{memhfixc}

\jvol{00}
\firstpage{1}
\lastpage{}
\rhauthor{S. E. Tumasz and J.-L. Thiffeault}

\usepackage{imakeidx}
\makeindex[name=subject,title=Subject index,columns=1]
\makeindex[name=authors,title=Author index,columns=1]

\newcommand{\mathnotation}[2]{\newcommand{#1}{\ensuremath{#2}}}
\mathnotation{\hrods}{h_{\text{rods}}}
\mathnotation{\s}{\sigma}

\usepackage{color}

\begin{document}

\dochead{Topological Fluid Dynamics II}
\title{Estimating topological entropy from the motion of stirring rods}
\author{Sarah E. Tumasz and Jean-Luc Thiffeault$^{\hbox{*}}$}
\affiliation{Department of Mathematics, University of Wisconsin --
  Madison, 480 Lincoln Drive, Madison, WI, 53706, USA}

\index[authors]{Tumasz, S.E.}
\index[authors]{Thiffeault, J.-L.}

\startabstract{\begin{abstract}

    Stirring a two-dimensional viscous fluid with rods is often an
    effective way to mix.  The topological features of periodic rod
    motions give a lower bound on the topological entropy of the
    induced flow map, since material lines must `catch' on the rods.
    But how good is this lower bound?  We present examples from
    numerical simulations and speculate on what affects the `gap'
    between the lower bound and the measured topological entropy.  The
    key is the sign of the rod motion's action on first homology of
    the orientation double cover of the punctured disk.

\smallskip

{\copyright}~2012 Published by Elsevier Ltd. Selection and/or peer-review under responsibility of K.\ Bajer, Y.\ Kimura, \& H.K.\ Moffatt.

\keywords{{\it Keywords:} fluid stirring ; topological entropy ; braids}
\end{abstract}}

\maketitle
\correspondingauthor{Corresponding author. Tel.: +1-608-263-4089 ;
  fax: +1-608-263-8891 .}
\email{jeanluc@math.wisc.edu}

\section{Introduction}
\label{sec:intro}

\index[subject]{topological!stirring}
\index[subject]{topological!entropy}
\index[subject]{surface dynamics}
\index[subject]{stirring!rods}
\index[subject]{stirring!topological}
\index[subject]{topological!stirring}
The paper of Boyland, Aref \& Stremler~\cite{Boyland2000} pioneered
\index[authors]{Boyland, P.L.}
\index[authors]{Aref, H.}
\index[authors]{Stremler, M.A.}
the study of two-dimensional rod-stirring devices using tools from
topological surface dynamics.  The central idea is that some rod
motions impose a minimal complexity to the fluid trajectories,
resulting in good mixing
\index[subject]{mixing}
in at least part of the domain.  Since then, many studies have
followed: these include several papers dealing directly with rod
motion~\cite{
% Rods
MattFinn2003,
Vikhansky2004,
MattFinn2006,
MattFinn2007,
Kobayashi2007,
Binder2008,
Thiffeault2008b,
Boyland2011,
MattFinn2011_silver%
}; various work on vortices, `ghost rods,' and almost-invariant sets~\cite{
% Ghost rods and AIS
Boyland2003,
Boyland2005,
Gouillart2006,
Stremler2007,
Thiffeault2009,
Binder2010,
Stremler2011%
};
papers on the topology of chaotic trajectories and random braids~\cite{
% Trajectories
Vikhansky2003,
Kin2005,
Thiffeault2005,
Allshouse2012,
Thiffeault2010,
Turner2011,
Puckett2012%
}; a paper on a extension to three dimensions using stationary rod
inserts~\cite{MattFinn2003b}; and a review~\cite{Thiffeault2006} and magazine
article~\cite{Thiffeault2009_catscradle}.

Throughout all this, there remains a vexing question, first raised by
Phil Boyland: if one studies the rod motion depicted in
\index[authors]{Boyland, P.L.}
Fig.~\ref{fig:braidlines_s1s-2}, which is
denoted~$\sigma_1\sigma_2^{-1}$ in terms of braid group generators,
\index[subject]{braid}
\index[subject]{braid!generators}
the growth rate of material lines \emph{in the fluid}
\index[subject]{material line}
is almost the same as that predicted by
the rod motion, which is a lower bound.  How do we explain such a
small discrepancy (or gap) between the lower bound and the measured
value?  Here we do not propose a full solution to this problem, but
instead offer some observations, based on numerical simulations, of
when the lower bound is and isn't sharp, what this correlates with,
and speculate on possible causes.  At the heart of the matter is
`secondary folding,'
\index[subject]{secondary folding}
or the observation that in some cases material lines fold a lot more
than is strictly required by the topology of the rod motion.  This
issue was explored in detail by the authors for toral linked twist
maps~\cite{Tumasz2012,Tumasz2012_thesis}.  Here we focus on physical
rod-stirring devices, also called rod mixers.

We can also interpret a small gap in terms of `taffy
pullers'~\cite{MattFinn2011_silver}.
\index[subject]{taffy pullers}
Ignore the fluid and consider a plastic `strap' wrapped around the
rods.  As the rods move, imagine the plastic strap can stretch, but
can never shrink.  The motions with a small gap described below will
lead to a plastic strap that never develops any slack throughout the
entire rod motion.  The question is to determine what properties of a
braid are needed to ensure this.

\section{Braid-based rod mixers}
\label{sec:braidmixers}

\begin{figure}
\captioncentertrue
\begin{center}
\subfigure[]{
\includegraphics[height=.2\textheight]{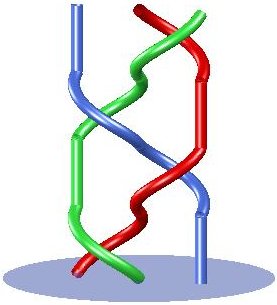}
\label{fig:braidlines_s1s-2}
}\hspace{4em}
\subfigure[]{
\includegraphics[height=.2\textheight]{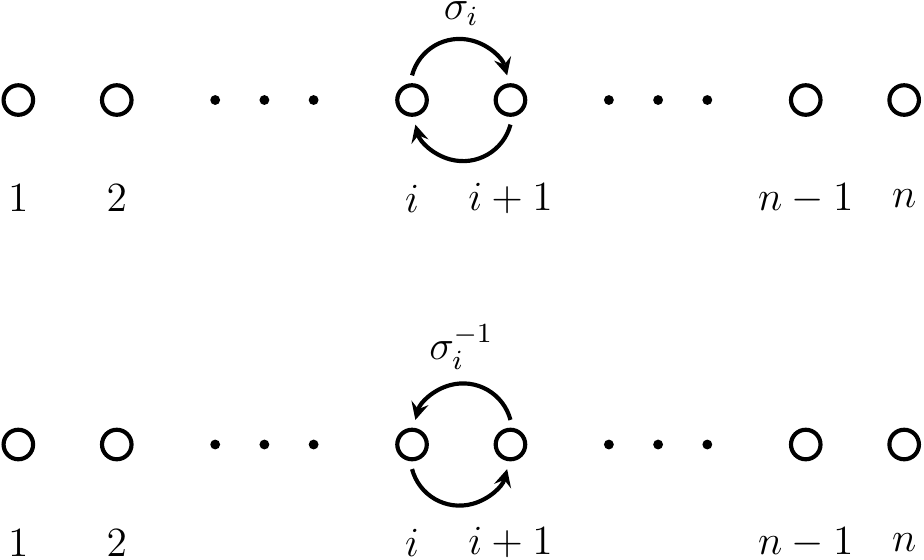}
\label{fig:braidgens}
}
\end{center}
\caption{(a) Motion of rods in a 3-rod mixer described by the braid
  $\sigma_1\sigma_2^{-1}$.  The vertical axis is time, and two full
  periods are shown. (b) The braid generator $\sigma_i$ (top) is a
  clockwise interchange of the $i$th and $(i+1)$th rods, with all
  other rods held fixed.  Its inverse, $\sigma_i^{-1}$ (bottom), is a
  counter-clockwise interchange of the same two rods.}
\end{figure}

We consider rod-stirring devices or mixers that are constructed such
that the rod motion is described by
\emph{braids}~\cite{Artin1947,Birman1975,Birman2005}.
\index[subject]{braid}
In these two-dimensional circular containers, the rods start along a
fixed horizontal line and move in accordance with \emph{braid
  generators}, $\sigma_i^{\pm 1}$, depicted in
Figure~\ref{fig:braidgens}.  For example, in a 3-rod mixer given by
the braid~$\sigma_1\sigma_2^{-1}$~\cite{Boyland2000}, first the two
leftmost rods move halfway around a circle in a clockwise direction
clockwise.  Immediately after that, the two rightmost rods move
halfway around a circle in a counter-clockwise direction (Figure
\ref{fig:braidlines_s1s-2}).  The circular paths are centred directly
between the two rods, and have diameter equal to the rod spacing.  The
speed of the rods is immaterial, since we are only considering Stokes
(slow viscous) flow.

More generally, we write the stirring motion for~$n$ rods as a braid
expressed as a sequence of generators, $\sigma_i$, $i=1,\ldots,n-1$.
Each generator represents the clockwise interchange of the $i$th and
$(i+1)$th strands or rods.  The inverse, $\sigma_i^{-1}$, is a
counter-clockwise interchange (Figure \ref{fig:braidgens}).  Note that
the strands are always numbered from left to right, so a given
subscript does not always refer to the same rod.  By having the rods
move in the same way as a specific braid, we can directly and
systematically compare the measured topological entropy in the fluid
system to the lower bound predicted by the braid (via the isotopy
class~\cite{Fathi1979,Thurston1988,Boyland1994}).
\index[subject]{isotopy}
\index[subject]{Thurston--Nielsen classification}
\index[authors]{Thurston, W.P.}
\index[authors]{Nielsen, J.}

\paragraph{Remark} There are different conventions in the literature:
In some papers $\sigma_i$ is defined as the counter-clockwise
interchange, which is the opposite of our definition.  There are also
differing conventions on composition order.  We will always write
generators from left to right -- that is, in the braid
$\sigma_1\sigma_2$, the $\sigma_1$ interchange occurs \emph{before}
the $\sigma_2$ interchange.

\paragraph{Remark} The lower bound on the entropy, based on the braid,
is independent of the specific details of the rod
motion.
\index[subject]{topological!entropy}
However, the measured flow entropy depends in general on the rod
radius, rotation, and how near the rods come to each other and to the
outer wall of the container during their motion.  In our simulations,
the rod radii are relatively small and we keep them from coming too
close to the wall to avoid extra growth of material lines due to image
effects.  Our simulations were performed with the computer program
\emph{Flop}, by Matthew D.\ Finn, Emmanuelle Gouillart, and J.-L.T.
\index[authors]{Finn, M.D.}
\index[authors]{Gouillart, E.}
The program is based on the complex-variable method described
in~\cite{MattFinn2003}.  We measure the flow topological entropy~$h$
from the growth rate of material lines in the
flow~\cite{Yomdin1987,Newhouse1988,Newhouse1993}.
\index[subject]{material line}

%%%%%%%%%%%%%%%%%%%%%%%%%%%%%%%%%%%%%%%%%%%%%%%%%%%%%%%%%%%%%%%%%%%%%%%%%%
\subsection{Three-rod mixers}
\label{sec:threerodmixers}

We start by looking at devices with three rods.  In particular, we
will focus on motions based on braids of the form
$\sigma_1^k\sigma_2^{-\ell}$.  When $k\ell>0$, we call the braid
\emph{counter-rotating}; when $k\ell<0$ we call it \emph{co-rotating}.
Braids of this form are \emph{pseudo-Anosov} if and only if
$|2+k\ell|>2$.  All counter-rotating braids are pseudo-Anosov, but
co-rotating braids are only pseudo-Anosov if $k\ell<-4$.
\index[subject]{pseudo-Anosov braid}
\index[subject]{braid!pseudo-Anosov}
Braids that are not pseudo-Anosov are \emph{finite order} or
\emph{reducible}, according to the Thurston--Nielsen classification
theorem~\cite{Fathi1979,Thurston1988,Boyland1994}.  We will not
encounter any reducible braids in this paper.
\index[subject]{braid!finite-order}
\index[subject]{braid!reducible}

\begin{figure}
\captioncentertrue
\begin{center}
\hfill
    \subfigure[$\sigma_1\sigma_2^{-1}$ (5 periods)]
{\includegraphics[width = .34\textwidth]{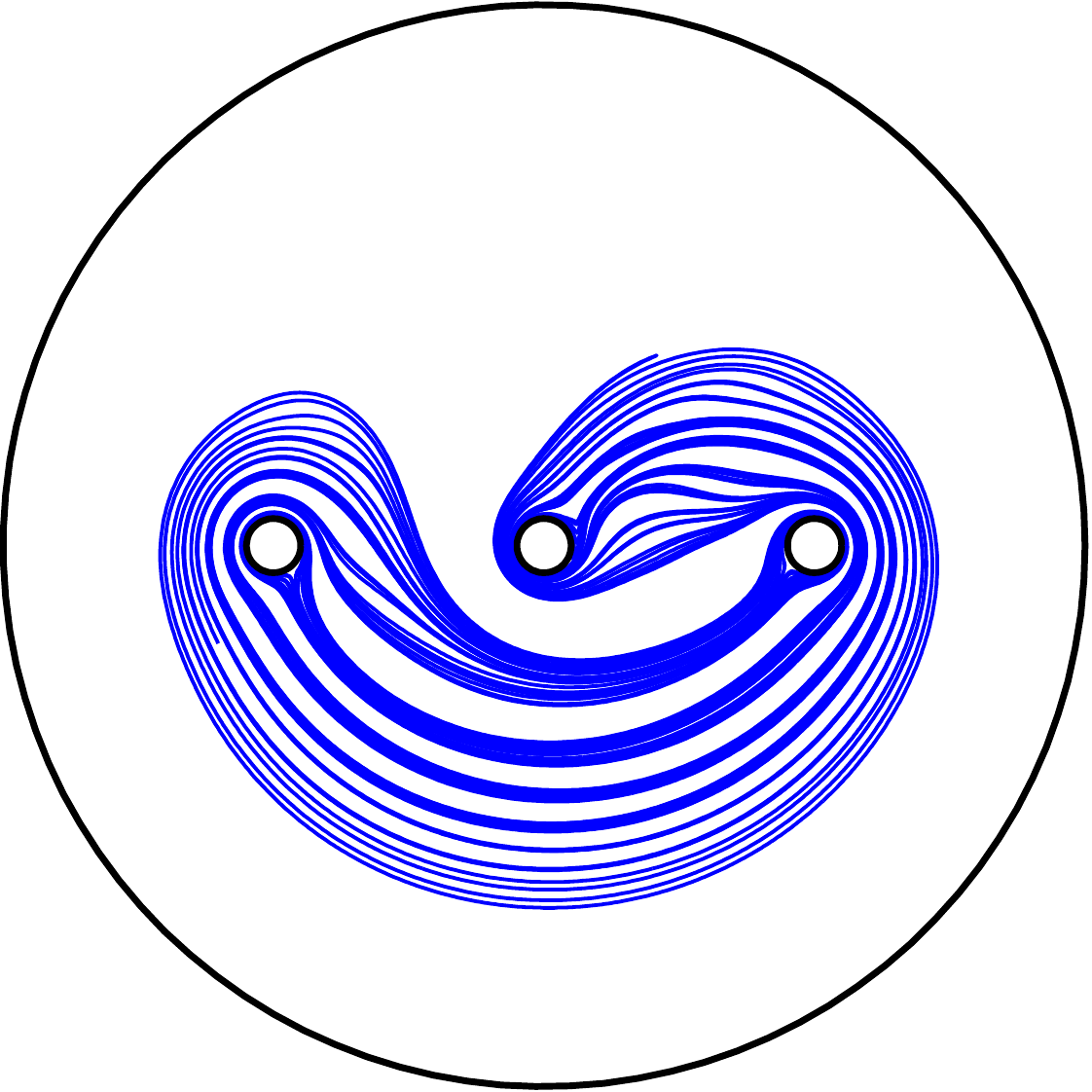}} \hfill
    \subfigure[$\sigma_1\sigma_2$ (9 periods)]
{\includegraphics[width = .34\textwidth]{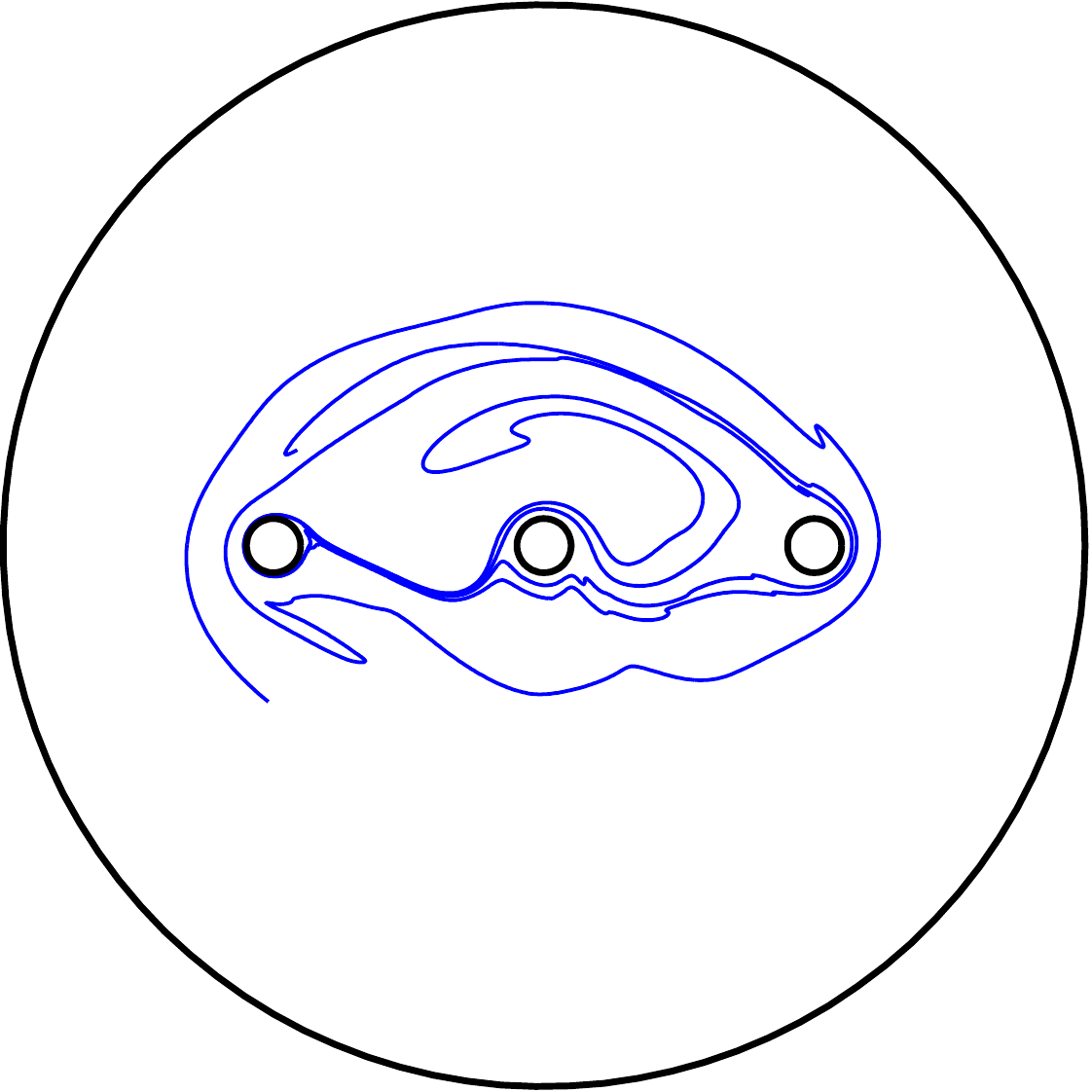}} \hfill \,

\hfill
      \subfigure[$\sigma_1\sigma_2^{-5}$ (3 periods)]
{\includegraphics[width = .34\textwidth]{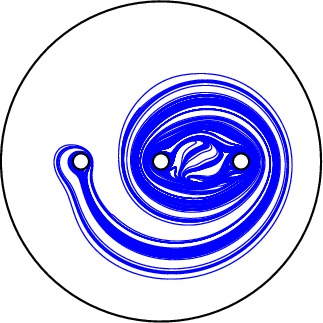}} \hfill
      \subfigure[$\sigma_1\sigma_2^{5}$ (3 periods)]
{\includegraphics[width = .34\textwidth]{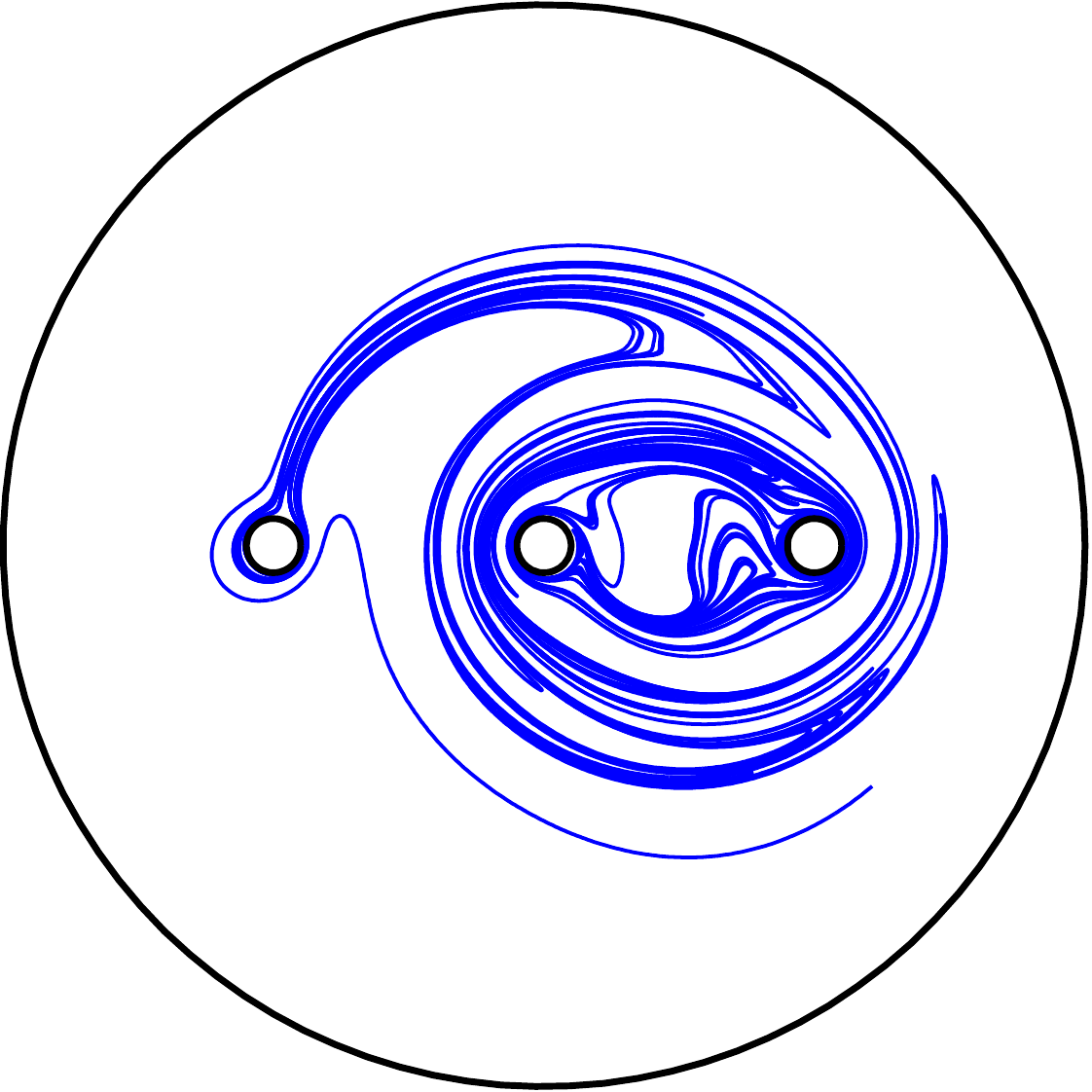}} \hfill \,

\hfill
    \subfigure[$\sigma_1^2\sigma_2^{-3}$ (3 periods)]
{\includegraphics[width = .34\textwidth]{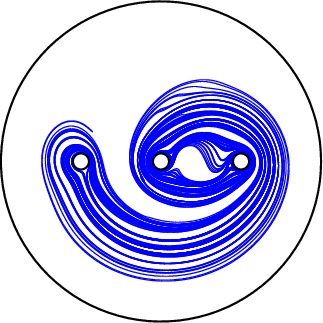}} \hfill
    \subfigure[$\sigma_1^2\sigma_2^{3}$ (3 periods)]
{\includegraphics[width = .34\textwidth]{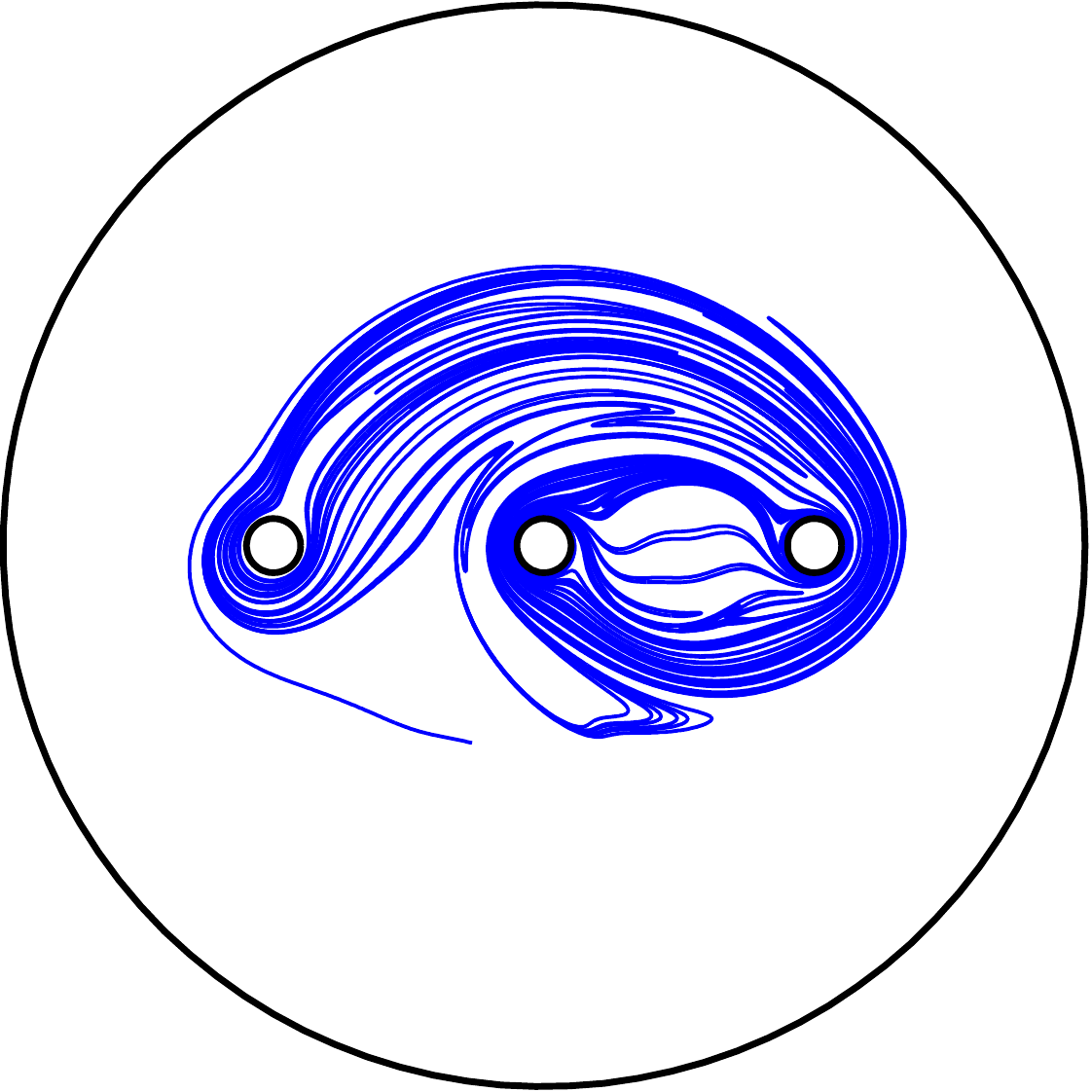}} \hfill \,
\end{center}
\caption{Material line patterns for several three-rod braid
  mixers. \index[subject]{material line}}
\label{fig:sigmas_threerod}
\end{figure}

Figure \ref{fig:sigmas_threerod} shows an iterated material line
\index[subject]{material line}
for several different braid mixers.
The three in the left column are counter-rotating, and the three in
the right column are co-rotating.
Of the six braid mixers shown in Figure~\ref{fig:sigmas_threerod},
five are pseudo-Anosov and one is not.  With only a quick glance, it
is not hard to guess that $\sigma_1\sigma_2$ is the odd one out -- in
comparison to the others, the material line
\index[subject]{material line}
in that device has hardly stretched at all, even after 9
periods, and pseudo-Anosov braids have an exponential line stretching
rate~\cite{Fathi1979,Thurston1988,Boyland1994}.
\index[subject]{pseudo-Anosov braid}
\index[subject]{braid!pseudo-Anosov}

However, despite the fact that the \emph{braid} $\sigma_1\sigma_2$ is
not pseudo-Anosov, we still measure a positive topological entropy for
the \emph{flow} in the mixing device.
\index[subject]{topological!entropy}
In fact, the braid mixers tend to fall into two categories: those
where the flow entropy~$h$ is close to the braid entropy~$\hrods$ (of
the order of~$10\%$ difference), and those where~$h$ is considerably
larger than $\hrods$ ($>25\%$ difference).  Table~\ref{tab:3braids}
shows, for several braids of the form $\sigma_1^k\sigma_2^{-\ell}$,
the measured topological entropy in the braid mixer ($h$) and the
lower bound obtained from the rod braid ($\hrods$).  The last column
gives the `gap' between the two values, expressed as a percentage
of~$h$.  The first set of braids is counter-rotating ($k\ell<0$); the
second set co-rotating ($k\ell>0$).  Note that the counter-rotating
mixers show a small gap, and the co-rotating ones have a much larger
gap.

\begin{figure}
\captioncentertrue
\begin{center}
\subfigure[]{%
\begin{minipage}[c][1.3\width]{.17\textwidth}%
\includegraphics[width=1\textwidth]{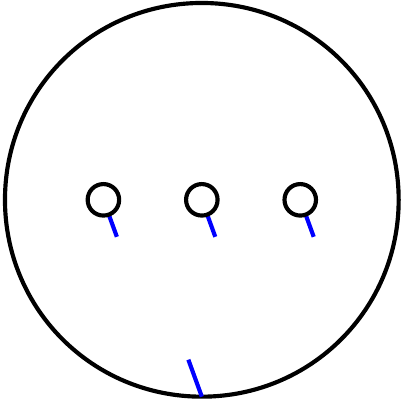}%
\end{minipage}%
}\hspace{2em}%
\subfigure[]{%
\begin{minipage}[c][1.3\width]{.17\textwidth}%
\includegraphics[width=1\textwidth]{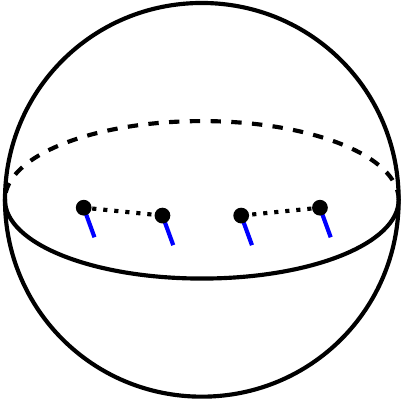}%
\end{minipage}%
}\hspace{2em}%
\subfigure[]{%
\begin{minipage}[c][1.3\width]{.17\textwidth}%
\includegraphics[width=1\textwidth]{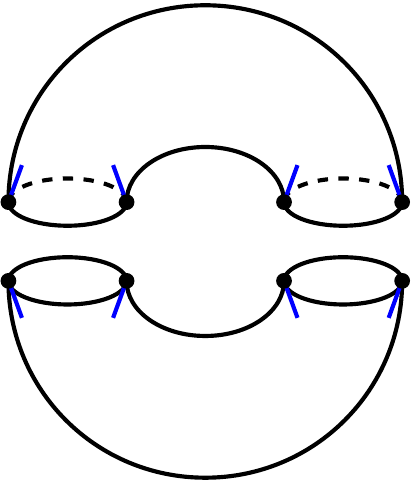}%
\end{minipage}%
}\hspace{2em}%
\subfigure[]{%
\begin{minipage}[c][1.3\width]{.17\textwidth}%
\includegraphics[width=1\textwidth]{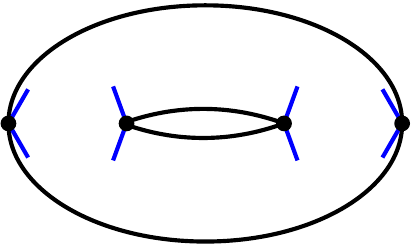}%
\end{minipage}%
}%
\end{center}
\caption{
  How to make the orientation double cover.
  (a) A disk with three rods, the small segments indicating
  pronged singularities.  (b) Shrink the disk's boundary and the
  rods to points, so the surface is a topological sphere.  Make two
  cuts between the rods and the boundary (dotted lines). (c) Glue a
  second copy of the same sphere along the cuts.  (d) The resulting
  surface is a torus, and the singularities now have two
  prongs (regular points).
  \index[subject]{orientation double cover}}
\label{fig:doublecover}
\end{figure}

The penultimate column of Table~\ref{tab:3braids} gives the sign of
the dominant eigenvalue (the one with the largest magnitude) of the
\index[subject]{braid!Burau representation}
\index[subject]{Burau representation}
Burau matrix representation of the braid.  The Burau
\index[authors]{Burau, W.}
representation~\cite{Burau1936,Birman1975,Fried1986,Kolev1989,BandBoyland2007}
arises from an action of the braid on first homology
\index[subject]{homology}
of a double cover of the punctured disk (actually a
$\mathbb{Z}$-cover, but we only use the double cover here).
Figure~\ref{fig:doublecover} depicts the construction of the double
cover for a disk with three rods.  Notice in Table~\ref{tab:3braids}
that for the pseudo-Anosov braids ($\hrods>0$), all the
counter-rotating cases have a negative Burau eigenvalue, while all the
co-rotating cases have a positive eigenvalue.  For the
non-pseudo-Anosov braids (i.e.\ those of finite order), the eigenvalue
\index[subject]{braid!finite-order}
\index[subject]{Burau representation}
\index[subject]{braid!Burau representation}
of the Burau matrix is always on the unit circle (complex), so we do
not record a sign.

For 3-braids, the logarithm of the spectral radius of the Burau matrix
agrees with the topological entropy of the braid.  For pseudo-Anosov
braids this largest eigenvalue is real but can be either positive or
negative.  A negative eigenvalue corresponds to a `flip' of the
homological generators at every application of the braid.  For toral
linked twist maps, this is associated with `kinks' in the material
lines,
\index[subject]{material line}
as shown in~\cite{Tumasz2012}.  These are what we call `secondary
folds,' as depicted for a fluid system in
Figure~\ref{fig:secondaryfolding} and discussed in
Section~\ref{sec:secondaryfolding}).
\index[subject]{secondary folding}
The conjecture is that these kinks lead to additional growth of
material lines, thus causing extra entropy above the lower bound.
However, this connection has not been yet rigorously
demonstrated.
\index[subject]{topological!entropy}

\begin{table}[h]
  \caption{Measured topological entropy vs.\ the lower bound for 3-rod
    braid mixers.  The sign listed is that of the dominant eigenvalue
    in the Burau matrix.}
\label{tab:3braids}
\vspace{.25em}
\tablefont
\begin{tabular*}{\hsize}{@{\extracolsep{\fill}}lllll@{}}%{lllll}
\hline
 braid           &  $h$    & $\hrods$ & Burau sign & gap \\
\hline
 $\s_1\s_2^{-1}$    &  0.992  &  0.9624  &  pos &  3.0\%  \\
 $\s_1\s_2^{-2}$    &  1.380  &  1.3170  &  pos &  4.6\%   \\
 $\s_1\s_2^{-3}$    &  1.714  &  1.5668  &  pos &  8.6\% \\
 $\s_1\s_2^{-4}$    &  2.048  &  1.7627  &  pos &  13.9\%  \\
 $\s_1\s_2^{-5}$    &  2.112  &  1.9248  &  pos &  8.9\% \\
 $\s_1^2\s_2^{-2}$  &  1.867  &  1.7627  &  pos &  5.6\%  \\
 $\s_1^2\s_2^{-3}$  &  2.244  &  2.0634  &  pos &  8.0\%  \\
 $\s_1^2\s_2^{-4}$  &  2.612  &  2.2924  &  pos &  12.2\%  \\
\hline
 $\s_1\s_2$         &  0.289  &  0       &      &  100\% \\
 $\s_1\s_2^2$       &  0.550  &  0       &      &  100\%  \\
 $\s_1\s_2^3$       &  1.109  &  0       &      &  100\%  \\
 $\s_1\s_2^4$       &  1.829  &  0       &      &  100\%  \\
 $\s_1\s_2^5$       &  1.611  &  0.9624  &  neg &  40.3\%  \\
 $\s_1^2\s_2^2$     &  1.328  &  0       &      &  100\%  \\
 $\s_1^2\s_2^3$     &  1.762  &  1.3170  &  neg &  25.3\%  \\
 $\s_1^2\s_2^4$     &  2.455  &  1.7627  &  neg &  28.2\%  \\
\hline
\end{tabular*}
\end{table}

%%%%%%%%%%%%%%%%%%%%%%%%%%%%%%%%%%%%%%%%%%%%%%%%%%%%%%%%%%%%%%%%%%%%%%%%%%
\subsection{Four-rod mixers}
\label{sec:fourrodmixers}

We now look at four-rod mixing devices.  With four rods, there is no
sense in classifying braids as counter- or co-rotating.  Instead, we
will focus on the sign of the dominant eigenvalue of the Burau matrix.
Since we have more than three rods, the dominant eigenvalue of the
Burau matrix is no longer guaranteed to give the topological entropy
of the braid -- it merely provides a lower
bound~\cite{Fried1986,Kolev1989}.  Band \&
\index[authors]{Band, G.}
\index[authors]{Boyland, P.L.}
Boyland~\cite{BandBoyland2007} showed the Burau eigenvalue gives the
exact topological entropy for a pseudo-Anosov braid if and only if the
corresponding foliation
\index[subject]{foliation}
\index[subject]{foliation!singularities in}
has odd-order singularities at all the
punctures, and any interior singularities are of even order.  One
consequence is that the Burau bound is always sharp for 3-braids, a
fact we used in Section~\ref{sec:threerodmixers} to compute the
entropy.
\index[subject]{topological!entropy}
\index[subject]{braid!Burau representation}
\index[subject]{Burau representation}
\index[subject]{pseudo-Anosov braid}
\index[subject]{braid!pseudo-Anosov}

\begin{figure}
\captioncentertrue
\begin{center}
\hfill
    \subfigure[$\sigma_1\sigma_2^{-1}\sigma_3\sigma_2^{-1}$ (3 periods)]
{\includegraphics[width = .34\textwidth]{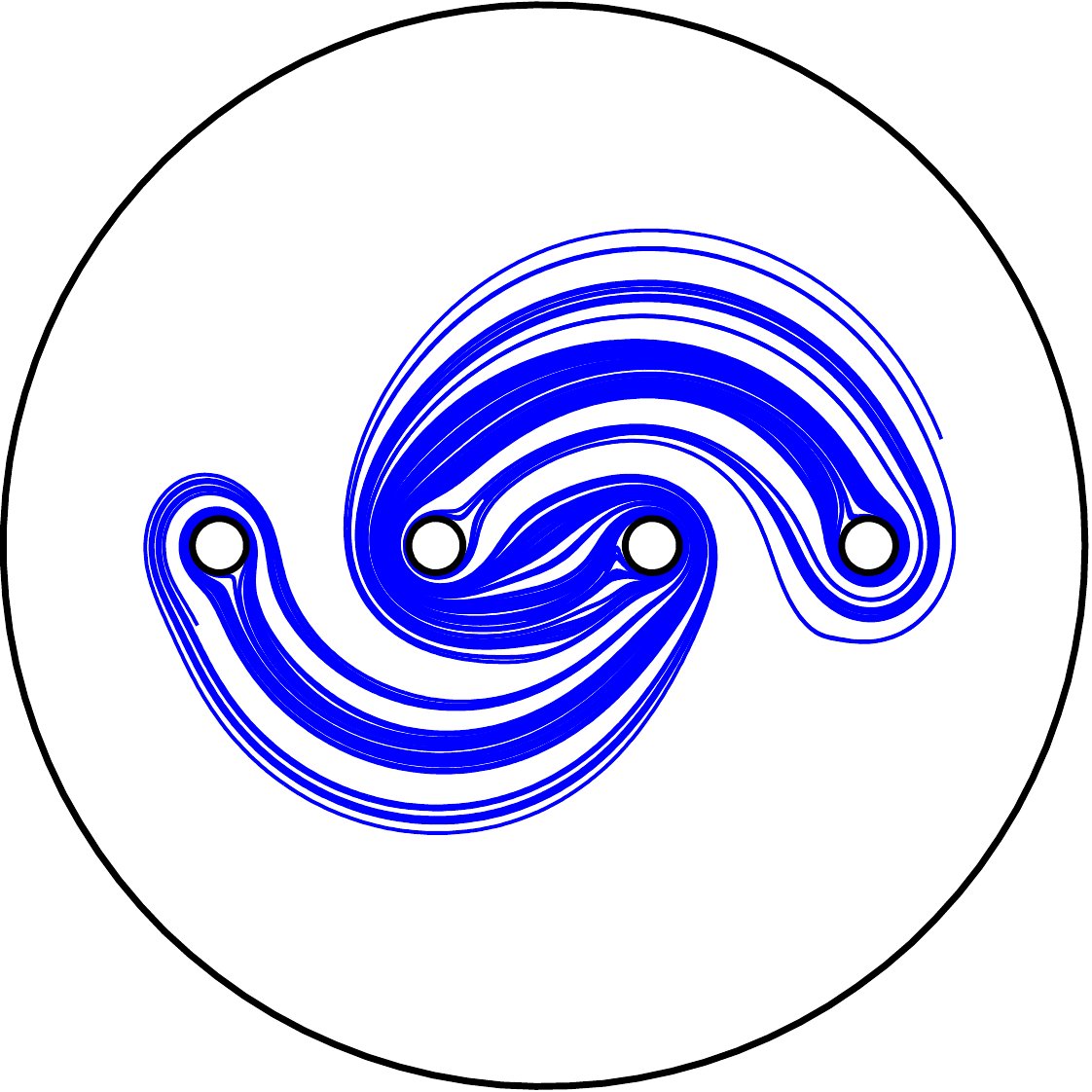}} \hfill
 \subfigure[$\sigma_1^2\sigma_2^{-4}\sigma_3^2$ (2 periods)]  %the Nicos braid
{\includegraphics[width = .34\textwidth]{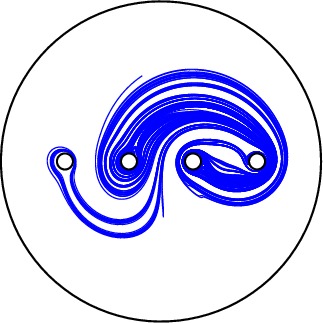}} \hfill \,

\hfill
    \subfigure[$\sigma_1\sigma_2\sigma_3^{-1}$ (6 periods)]
{\includegraphics[width = .34\textwidth]{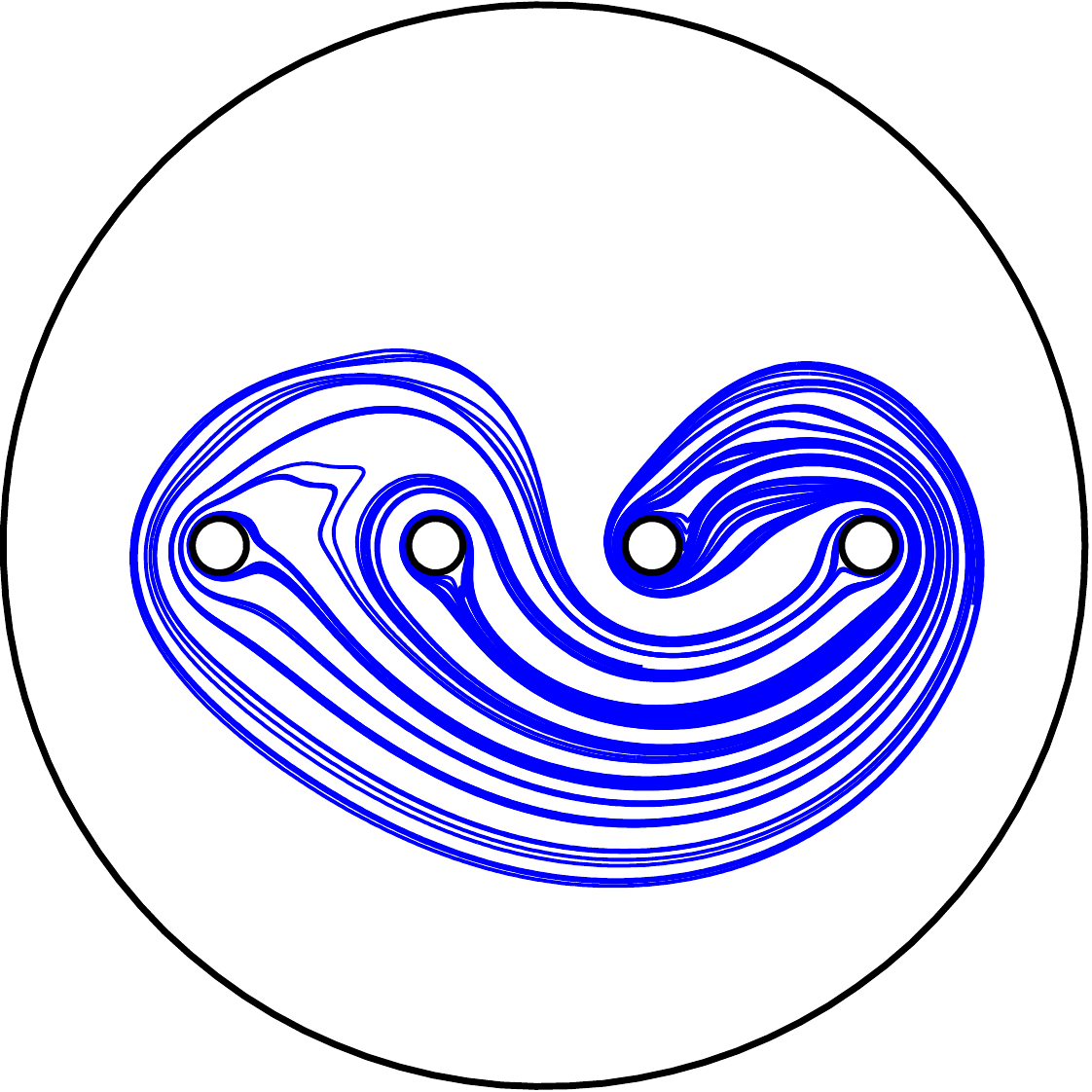}} \hfill
         \subfigure[$\sigma_1\sigma_2\sigma_3^{-1}\sigma_2^{-1}$ (5 periods)]
{\includegraphics[width = .34\textwidth]{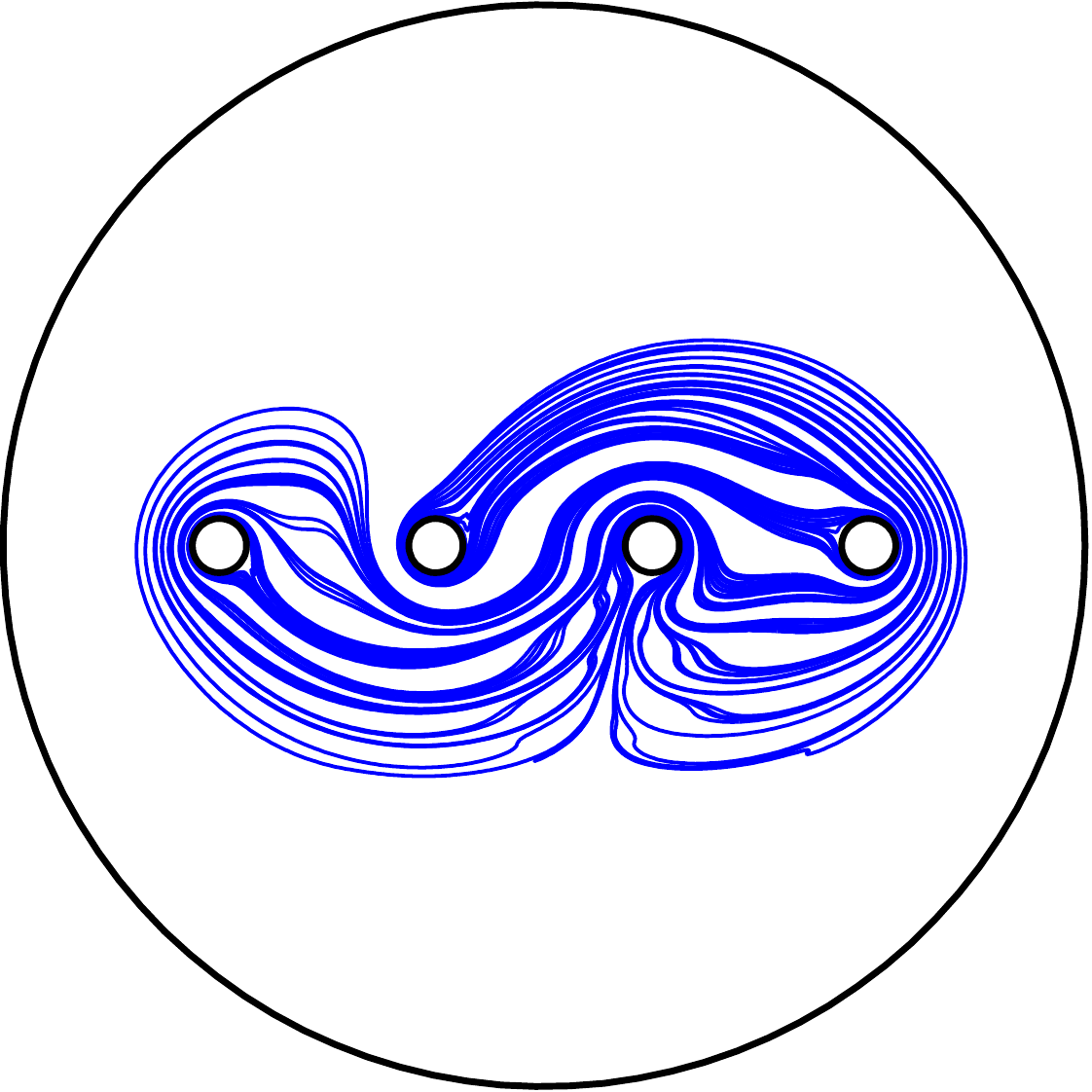}} \hfill \,

\hfill
    \subfigure[$\sigma_1\sigma_2^{2}\sigma_3^2$ (3 periods)]
{\includegraphics[width = .34\textwidth]{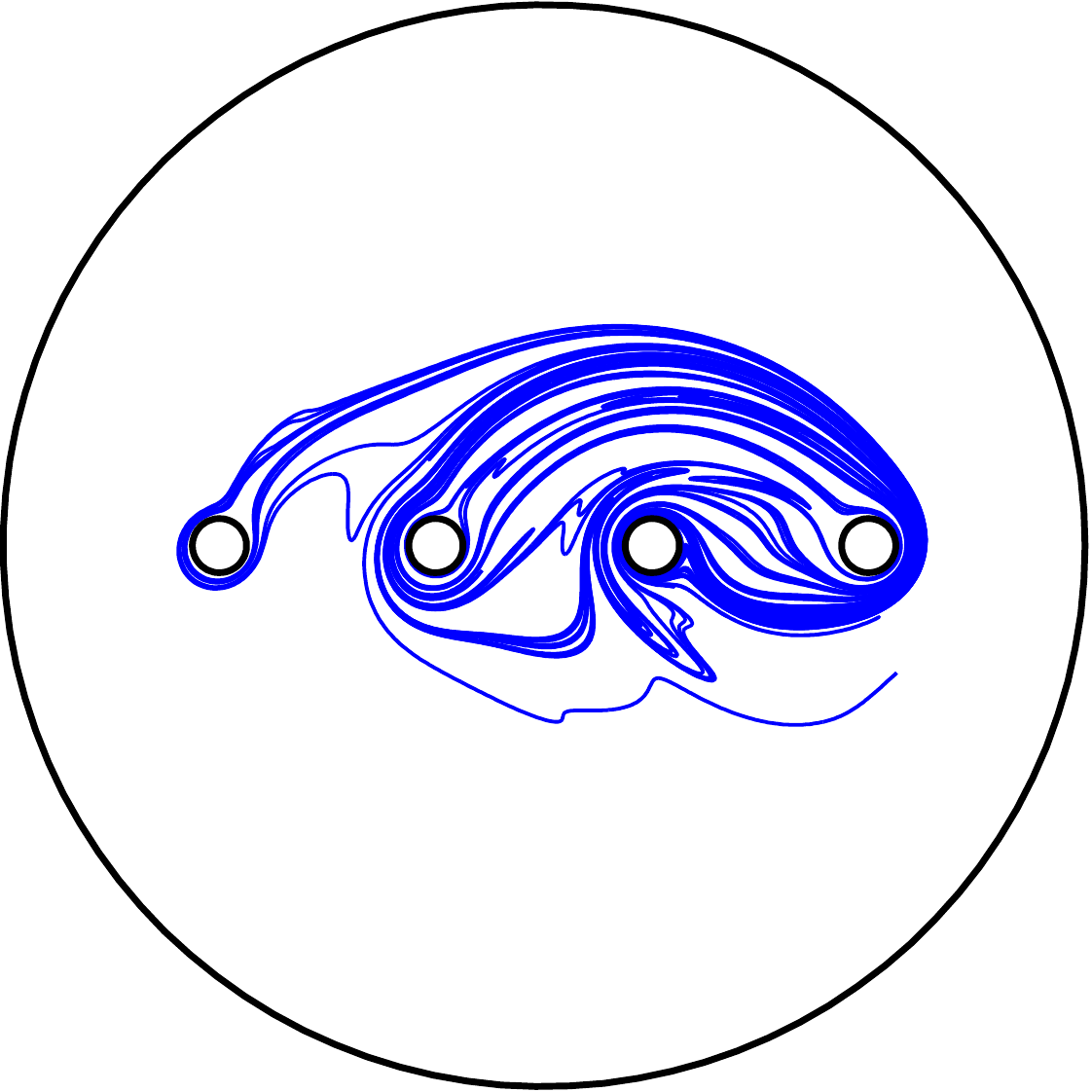}} \hfill
      \subfigure[$\sigma_1\sigma_2^{3}\sigma_3\sigma_2$ (4 periods)]
{\includegraphics[width = .34\textwidth]{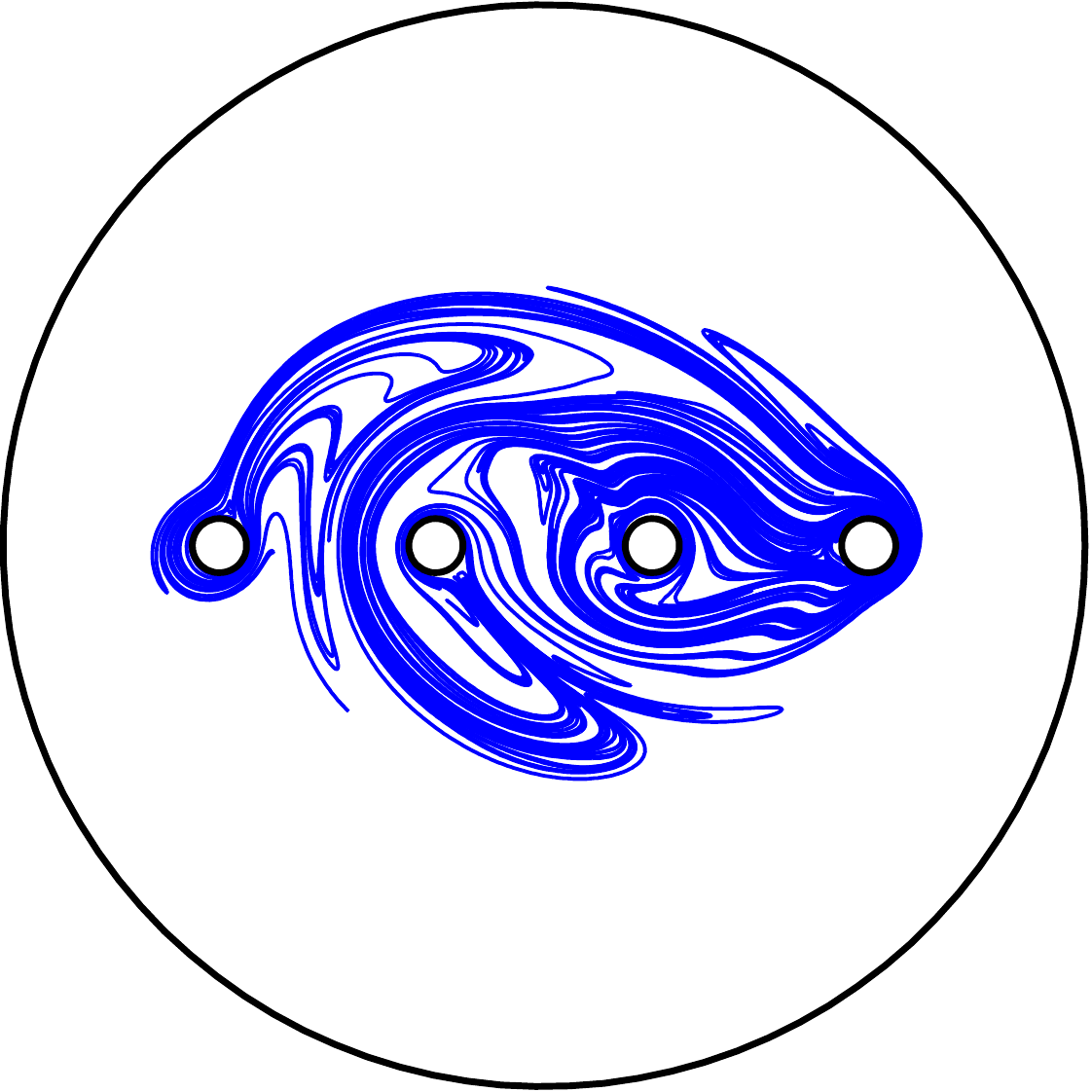}} \hfill \,
\end{center}
\caption{Material line patterns for several four-rod braid
  mixers. \index[subject]{material line}}
\label{fig:sigmas_fourrod}
\end{figure}

Figure \ref{fig:sigmas_fourrod} shows material line
\index[subject]{material line}
patterns for some four-rod mixers.
Table \ref{tab:4braids} lists the braid, the measured topological
entropy ($h$), the topological entropy of the braid ($\hrods$) given
by the Bestvina--Handel algorithm~\cite{Bestvina1995,HallTrain}, the
Burau bound, the sign of the dominant eigenvalue in the Burau matrix,
and size of the gap between the two topological entropy values.
Observe that again the braids with a positive Burau eigenvalue have
small gaps in entropy -- less than~$2\%$.  We will discuss the sources
of discrepancy in the next section.
\index[subject]{braid!Burau representation}
\index[subject]{Burau representation}

\begin{table}[h]
  \caption{Measured topological entropy vs.\ the lower bounds for 4-rod
    braid mixers.  The sign listed is that of the dominant
    eigenvalue in the Burau matrix.  The braids in the last set have
    vanishing Burau bound, so no sign is ascribed, except for the last
    braid (marked with~$^\dagger$) where we found the action on
    homology to be positive (Section~\ref{sec:disc}).}
\label{tab:4braids}
\vspace{.25em}
\tablefont
\begin{tabular*}{\hsize}{@{\extracolsep{\fill}}llllll@{}}%{lllll}
\hline
braid                   & $h$ & $\hrods$ & Burau bound & Burau sign & gap \\
\hline
$\s_1\s_2^{-1}\s_3\s_2^{-1}$      & 1.949 & 1.92485 & 1.92485 & pos &  1.2\% \\
$\s_1\s_2\s_3^{-1}\s_2^{-1}$      & 0.970 & 0.96242 & 0.96242 & pos &  0.8\%  \\
$\s_1\s_3\s_2^{-1}$               & 1.319 & 1.31696 & 1.31696 & pos &  0.2\%   \\
$\s_1^2\s_2^{-2}\s_3^2\s_2^{-2}$  & 3.559 & 3.52549 & 3.52549 & pos &  0.9\%  \\
$\s_1^2\s_2^{-4}\s_3^2$           & 2.940 & 2.88727 & 2.88727 & pos &  1.8\%   \\
$\s_1^2\s_2^{-2}\s_3^2$           & 2.303 & 2.29243 & 2.29243 & pos &  0.5\%   \\
\hline
$\s_1\s_2^2\s_3^2$                & 1.562 & 1.31696 & 1.31696 & neg &  15.7\%  \\
$\s_1^2\s_2^{-1}\s_3\s_2^2$       & 1.914 & 1.56686 & 1.56686 & neg &  18.1\%  \\
$\s_1\s_2^3\s_3\s_2$              & 1.270 & 0.96242 & 0.96242 & neg &  24.2\%  \\
$\s_1^2\s_3^{2}\s_2^2$            & 1.903 & 1.76275 & 1.76275 & neg &  7.4\%  \\
\hline
$\s_1\s_2\s_3\s_2$                & 0.509 & 0       & 0       &     &  100\%  \\
$\s_1\s_2^{-1}\s_3^{-1}\s_2^{-1}$ & 1.065 & 0.96242 & 0       &     &  9.6\%  \\
$\s_1\s_3\s_2$                    & 0.275 & 0       & 0       &     &  100\%   \\
$\s_1\s_2\s_3^{-1}$               & 0.837 & 0.83144 & 0       & pos$^\dagger$    &  0.7\%  \\
\hline
\end{tabular*}
\end{table}

%%%%%%%%%%%%%%%%%%%%%%%%%%%%%%%%%%%%%%%%%%%%%%%%%%%%%%%%%%%%%%%%%%%%%%%%%%
\section{Explaining the gap}

Our ultimate goal is to predict when the lower bound from the rod
motion is close to the measured topological entropy (small gap), and
when it is not (large gap).  Furthermore, we wish to understand what
causes a large gap, that is: what is it about the flow that creates
more topological entropy?  The easier question to answer, at least
partially, is why the lower bound fails.  We will address this first.
Then we will attempt to explain why it
happens.
\index[subject]{topological!entropy}

%%%%%%%%%%%%%%%%%%%%%%%%%%%%%%%%%%%%%%%%%%%%%%%%%%%%%%%%%%%%%%%%%%%%%%%%%%
\subsection{Why there is a gap -- secondary folding}
\label{sec:secondaryfolding}

\begin{figure}
\captioncentertrue
\begin{center}
\hfill
\subfigure[Folding around a rod]{\label{fig:secondaryfolding_1}\includegraphics[width=.3\textwidth]{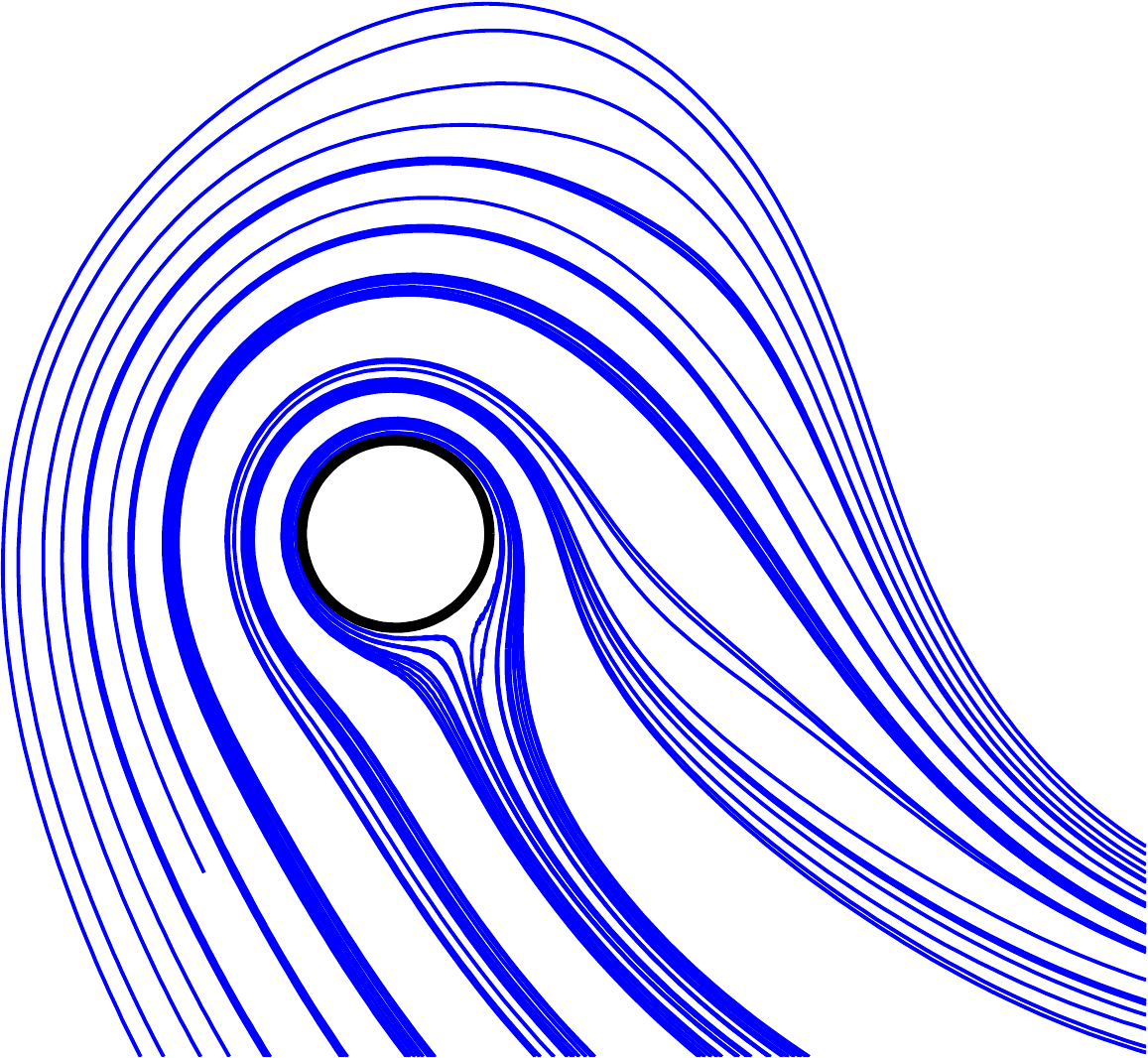}}
\hfill
\subfigure[Secondary Folding]{\label{fig:secondaryfolding_2}\includegraphics[width=.3\textwidth]{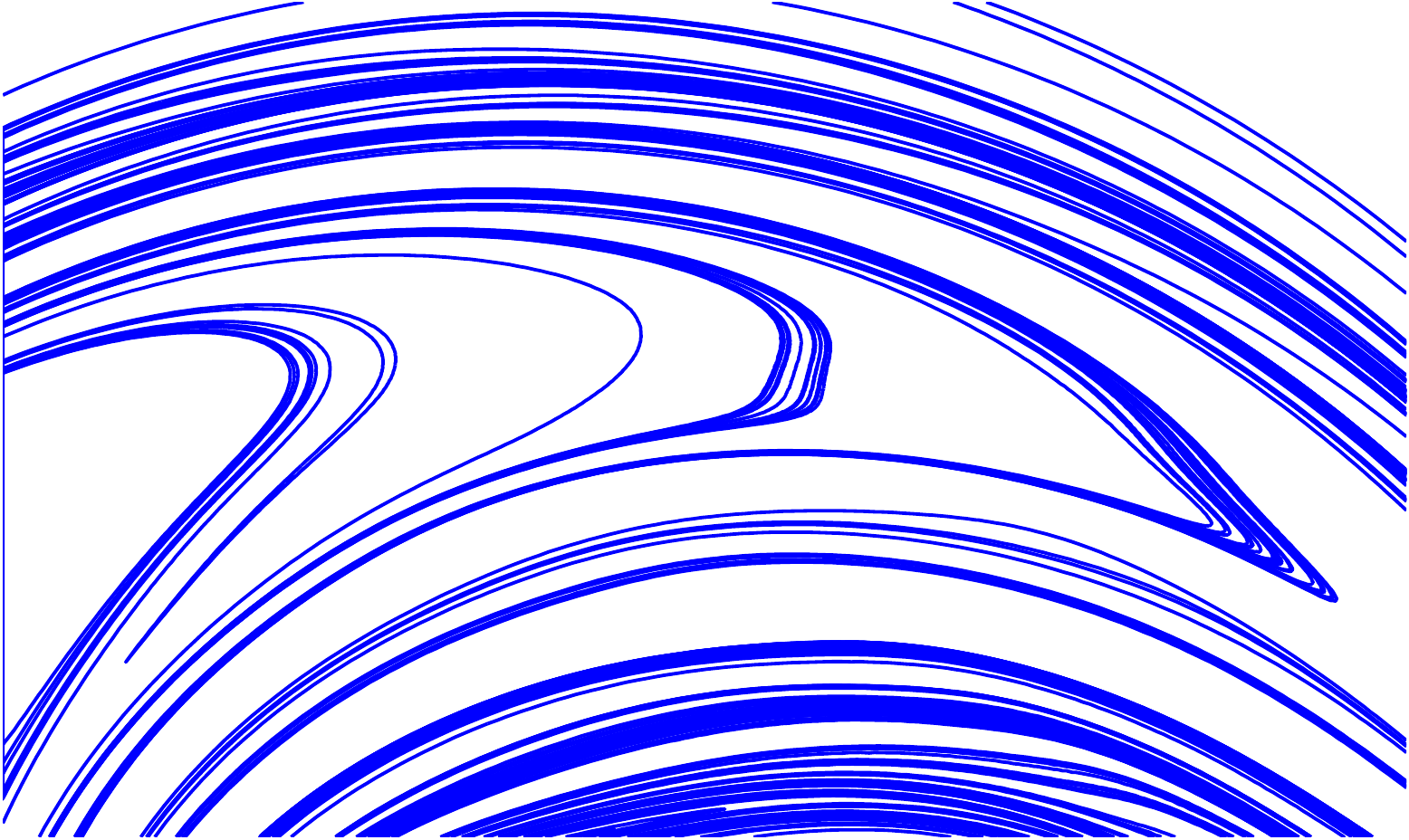}}
\hfill \,
\end{center}
\caption{(a) Folding caused by a rod moving through the fluid and
  dragging along the material lines. \index[subject]{material line}
  (b) Secondary folding of the material line.  This is not directly
  associated with folding around a rod and can be `pulled tight' via
  homotopy. \index[subject]{secondary folding}}
\label{fig:secondaryfolding}
\end{figure}

\index[subject]{secondary folding}
\index[subject]{isotopy}
Recall that the lower bound on entropy arises from the braid giving
the rod motion: this braid labels the isotopy class of the period-1
map.  Since the pseudo-Anosov representative of the isotopy class is
the `simplest' map in the class (the one with the lowest entropy), the
isotopy from the flow to the pseudo-Anosov representative has the
effect of pulling tight the material lines.
\index[subject]{material line}
In order for the flow to have a higher topological entropy, there must
be some part of the material line pattern that is not already
pulled-tight.  In other words, there must be some extra folding that
is not directly due to the rods.  We call this \emph{secondary
  folding}~\cite{Tumasz2012,Tumasz2012_thesis}.
Figure~\ref{fig:secondaryfolding_1} shows an example of folding due to
a rod, and Figure~\ref{fig:secondaryfolding_2} shows secondary
folding, which is not associated with a rod and could be removed by
pulling tight.
\index[subject]{topological!entropy}

Having a few extra folds is not necessarily enough to cause higher
topological entropy.  Recall that for 2D systems topological entropy
is related to the exponential stretching rate of material
lines~\cite{Yomdin1987,Newhouse1988,Newhouse1993}.  The extra folds
must cause a higher line growth rate in order to affect the
topological entropy.
\index[subject]{topological!entropy}
\index[subject]{material line}

Looking back at the three-braid mixers shown in
Figure~\ref{fig:sigmas_threerod}, there is visible secondary folding
\index[subject]{secondary folding}
in the $\sigma_1\sigma_2^5$ and $\sigma_1^2\sigma_2^3$ mixers.  From
Table~\ref{tab:3braids}, we see that these had gaps of $40.3\%$ and
$25.3\%$ respectively.  In contrast, the $\sigma_1\sigma_2^{-1}$,
$\sigma_1\sigma_2^{-5}$, and $\sigma_1^2\sigma_2^{-3}$ mixers have no
visible secondary folding, and have gaps of $3.0\%$, $8.9\%$, and
$8.0\%$ respectively.  The same can be seen in the four-rod braid
mixers of Figure \ref{fig:sigmas_fourrod}.  There is visible secondary
folding in the $\sigma_1\sigma_2^2\sigma_3^2$ and
$\sigma_1\sigma_2^3\sigma_3\sigma_2$ mixers, and these have gaps of
$15.7\%$ and $24.2\%$ respectively.  In comparison, for the
$\sigma_1\sigma_2^{-1}\sigma_3\sigma_2^{-1}$,
$\sigma_1^2\sigma_2^{-4}\sigma_3^2$, $\sigma_1\sigma_2\sigma_3^{-1}$,
and $\sigma_1\sigma_2\sigma_3^{-1}\sigma_2^{-1}$ mixers there is no
visible secondary folding, with gaps of $1.2\%$, $1.8\%$, $0.7\%$, and
$0.8\%$ respectively.

%%%%%%%%%%%%%%%%%%%%%%%%%%%%%%%%%%%%%%%%%%%%%%%%%%%%%%%%%%%%%%%%%%%%%%%%%%
\subsection{When there is a gap -- negative eigenvalues}

\index[subject]{topological!entropy}
We would now like to predict \emph{when} we can expect a gap between
the measured topological entropy and the lower bound given by the
braid.  From the data presented, it is tempting to say that mixers
with a braid whose Burau matrix has a negative dominant eigenvalue
have a large gap, while those with positive eigenvalues have a small
gap.  However, this says nothing about braids for which the Burau
bound is zero.  Furthermore, the data does not include any braids
whose Burau bound is non-zero, but also not equal to the topological
entropy of the braid (because of odd interior singularities in the
foliation).  We discuss why we didn't include such braids in
\index[subject]{foliation}
Section~\ref{sec:disc}.  However, it is clear at this point that the
sharpness is closely correlated with the sign of the action on first
homology
\index[subject]{homology}
of the orientation double cover,
\index[subject]{orientation double cover}
as given by the Burau representation
\index[subject]{braid!Burau representation}
\index[subject]{Burau representation}
in most cases examined here.

%\set{Try to do examples where Burau is not sharp and not zero. 1, 1,
%  1, 2, -3, 2 is such a braid.}

\section{Discussion}
\label{sec:disc}

In summary, we have exhibited a number of examples of braid-based rod
mixers.  These fall in two categories: those for which the rod motion
is a good predictor of the flow entropy, and those for which it
isn't.  For both three- and four-rod systems, the sign of the Burau
eigenvalue correlates well with the two cases: a positive eigenvalue
usually means that the bound is sharp.

When the Burau entropy is not sharp, the relevant quantity is the sign
of the action of the braid on homology lifted to the orientation
double cover.
\index[subject]{orientation double cover}
\index[subject]{homology}
When the sign is negative, then the entire homological chain must
`flip' which each action of the braid.  The conjecture is that this
flip causes secondary folding
\index[subject]{secondary folding}
by promoting `slack' in the material lines.
\index[subject]{material line}
This is evident when examining toral linked twist
maps~\cite{Tumasz2012,Tumasz2012_thesis}.  Unfortunately, this cannot
be the whole story, since repeating the rod motion twice will always
make the homological eigenvalue positive, but will clearly not make
the lower bound any better.
\index[subject]{topological!entropy}
\index[subject]{braid!Burau representation}
\index[subject]{Burau representation}

Why is the orientation double cover important?  The foliations
\index[subject]{foliation}
obtained on disks are always non-orientable, due to the odd-pronged
singularities at the rods.  The orientation double cover turns the
disk foliation into an orientable foliation on a closed surface of
some genus (a torus in Figure~\ref{fig:doublecover}).  It is then easy
to compute the topological entropy, since the linear action on
homology gives the entropy for the case of orientable foliations.
\index[subject]{foliation}
\index[subject]{foliation!orientable}
However, in order to construct the orientation double cover we need to
know \emph{a priori} the odd-pronged singularities associated with a
braid's isotopy class.
\index[subject]{isotopy}

In general we should be able to ascribe a homological sign even for
braids that are not Burau-sharp.  This is easy to do when a
pseudo-Anosov is given in terms of Dehn twists on the double
cover~\cite{LanneauThiffeault2011}, but is not so straightforward when
starting from braids on the disk; this is a future challenge.  For the
braid~$\sigma_1\sigma_2\sigma_3^{-1}$ in Table~\ref{tab:4braids}, we
were able to determine that the sign is positive by puncturing at the
3-pronged singularity and computing the Burau action of the resulting
5-braid ($(\sigma_1\sigma_2\sigma_1)\sigma_3\sigma_4^{-1}$).  Note
that both homological signs can always be realised, since the braids
giving rise to different signs are related by the deck transformation
(involution) of the double cover~\cite{LanneauThiffeault2011_braids}.
\index[subject]{topological!entropy}
\index[subject]{braid!Burau representation}
\index[subject]{Burau representation}
\index[subject]{pseudo-Anosov braid}
\index[subject]{braid!pseudo-Anosov}

\section*{Acknowledgements}

\index[authors]{Boyland, P.L.}
The authors thank Phil Boyland for his patient help.  J-LT is grateful
for the hospitality of the Isaac Newton Institute for Mathematical
Sciences in Cambridge, UK.  This work was funded by the Division of
Mathematical Sciences of the US National Science Foundation, under
grant DMS-0806821.

\bibliographystyle{vancouver}

\bibliography{bib/journals_abbrev,bib/articles}

\begin{thebibliography}{10}

\bibitem{Boyland2000}
Boyland PL, Aref H, Stremler MA.
\newblock Topological fluid mechanics of stirring.
\newblock J Fluid Mech. 2000;403:277--304.

\bibitem{MattFinn2003}
Finn MD, Cox SM, Byrne HM.
\newblock Topological chaos in inviscid and viscous mixers.
\newblock J Fluid Mech. 2003;493:345--361.

\bibitem{Vikhansky2004}
Vikhansky A.
\newblock Simulation of Topological Chaos in Laminar Flows.
\newblock Chaos. 2004 Mar;14(1):14--22.

\bibitem{MattFinn2006}
Finn MD, Thiffeault JL, Gouillart E.
\newblock Topological Chaos in Spatially Periodic Mixers.
\newblock Physica D. 2006 Sep;221(1):92--100.

\bibitem{MattFinn2007}
Finn MD, Thiffeault JL.
\newblock Topological Entropy of Braids on the Torus.
\newblock SIAM J Appl Dyn Sys. 2007;6:79--98.

\bibitem{Kobayashi2007}
Kobayashi T, Umeda S.
\newblock Realizing pseudo-{A}nosov egg beaters with simple mecanisms.
\newblock In: Proceedings of the International Workshop on Knot Theory for
  Scientific Objects, Osaka, Japan. Osaka Municipal Universities Press; 2007.
  p. 97--109.

\bibitem{Binder2008}
Binder BJ, Cox SM.
\newblock A Mixer Design for the Pigtail Braid.
\newblock Fluid Dyn Res. 2008;40:34--44.

\bibitem{Thiffeault2008b}
Thiffeault JL, Finn MD, Gouillart E, Hall T.
\newblock Topology of Chaotic Mixing Patterns.
\newblock Chaos. 2008 Sep;18:033123.

\bibitem{Boyland2011}
Boyland PL, Harrington J.
\newblock The entropy efficiency of point-push mapping classes on the punctured
  disk.
\newblock Algeb Geom Topology. 2011;11(4):2265--2296.

\bibitem{MattFinn2011_silver}
Finn MD, Thiffeault JL.
\newblock Topological optimisation of rod-stirring devices.
\newblock SIAM Rev. 2011 Dec;53(4):723--743.

\bibitem{Boyland2003}
Boyland PL, Stremler MA, Aref H.
\newblock Topological fluid mechanics of point vortex motions.
\newblock Physica D. 2003;175:69--95.

\bibitem{Boyland2005}
Boyland PL.
\newblock Dynamics of two-dimensional time-periodic {E}uler fluid flows.
\newblock Topology Appl. 2005;152:87--106.

\bibitem{Gouillart2006}
Gouillart E, Finn MD, Thiffeault JL.
\newblock Topological Mixing with Ghost Rods.
\newblock Phys Rev E. 2006;73:036311.

\bibitem{Stremler2007}
Stremler MA, Chen J.
\newblock Generating topological chaos in lid-driven cavity flow.
\newblock Phys Fluids. 2007;19:103602.

\bibitem{Thiffeault2009}
Thiffeault JL, Gouillart E, Finn MD.
\newblock The Size of Ghost Rods.
\newblock In: Cortelezzi L, Mezi\'{c} I, editors. Analysis and Control of
  Mixing with Applications to Micro and Macro Flow Processes. vol. 510 of
  {CISM} International Centre for Mechanical Sciences. Vienna: Springer; 2009.
  p. 339--350.

\bibitem{Binder2010}
Binder BJ.
\newblock Ghost rods adopting the role of withdrawn baffles in batch mixer
  designs.
\newblock Phys Lett A. 2010;374:3483--3486.

\bibitem{Stremler2011}
Stremler MA, Ross SD, Grover P, Kumar P.
\newblock Topological Chaos and Periodic Braiding of Almost-Cyclic Sets.
\newblock Phys Rev Lett. 2011;106:114101.

\bibitem{Vikhansky2003}
Vikhansky A.
\newblock Chaotic advection of finite-size bodies in a cavity flow.
\newblock Phys Fluids. 2003 Jul;15(7):1830--1836.

\bibitem{Kin2005}
Kin E, Sakajo T.
\newblock Efficient topological chaos embedded in the blinking vortex system.
\newblock Chaos. 2005;15(2):023111.

\bibitem{Thiffeault2005}
Thiffeault JL.
\newblock Measuring Topological Chaos.
\newblock Phys Rev Lett. 2005 Mar;94(8):084502.

\bibitem{Allshouse2012}
Allshouse MR, Thiffeault JL.
\newblock Detecting coherent structures using braids.
\newblock Physica D. 2012 Jan;241(2):95--105.

\bibitem{Thiffeault2010}
Thiffeault JL.
\newblock Braids of entangled particle trajectories.
\newblock Chaos. 2010 Jan;20:017516.

\bibitem{Turner2011}
Turner MR, Berger MA.
\newblock A study of mixing in coherent vortices using braiding factors.
\newblock Fluid Dyn Res. 2011 Jun;43(3):035501.

\bibitem{Puckett2012}
Puckett JG, Lechenault F, Daniels KE, Thiffeault JL.
\newblock Trajectory entanglement in dense granular materials.
\newblock Journal of Statistical Mechanics: Theory and Experiment. 2012
  Jun;2012(6):P06008.
\newblock Available from:
  \url{http://iopscience.iop.org/1742-5468/2012/06/P06008}.

\bibitem{MattFinn2003b}
Finn MD, Cox SM, Byrne HM.
\newblock Chaotic advection in a braided pipe mixer.
\newblock Phys Fluids. 2003 Nov;15(11):L77--L80.

\bibitem{Thiffeault2006}
Thiffeault JL, Finn MD.
\newblock Topology, Braids, and Mixing in Fluids.
\newblock Phil Trans R Soc Lond A. 2006 Dec;364:3251--3266.

\bibitem{Thiffeault2009_catscradle}
Thiffeault JL, Lanneau E, Matz SE. The cat's cradle, stirring, and topological
  complexity; 2009.
\newblock Http://www.dynamicalsystems.org/ma/ma/display?item=292.
\newblock Dynamical Systems Magazine.

\bibitem{Tumasz2012}
Tumasz SE, Thiffeault JL. Topological entropy and secondary folding; 2012.
\newblock {http://arXiv.org/abs/1204.6730}.

\bibitem{Tumasz2012_thesis}
Tumasz SE.
\newblock Topological Stirring.
\newblock University of Wisconsin -- Madison. Madison, WI; 2012.

\bibitem{Artin1947}
Artin E.
\newblock Theory of Braids.
\newblock Ann Math. 1947 Jan;48(1):101--126.

\bibitem{Birman1975}
Birman JS.
\newblock Braids, Links, and Mapping Class Groups.
\newblock No.~82 in Annals of Mathematics Studies. Princeton, NJ: Princeton
  University Press; 1975.

\bibitem{Birman2005}
Birman JS, Brendle TE.
\newblock Braids: {A} Survey.
\newblock In: Menasco W, Thistlethwaite M, editors. Handbook of Knot Theory.
  Amsterdam: Elsevier; 2005. p. 19--104.
\newblock Available at {\tt http://arXiv.org/abs/math.GT/0409205}.

\bibitem{Fathi1979}
Fathi A, Laundenbach F, Po\'{e}naru V.
\newblock Travaux de {T}hurston sur les surfaces.
\newblock Ast\'{e}risque. 1979;66-67:1--284.

\bibitem{Thurston1988}
Thurston WP.
\newblock On the geometry and dynamics of diffeomorphisms of surfaces.
\newblock Bull Am Math Soc. 1988;19:417--431.

\bibitem{Boyland1994}
Boyland PL.
\newblock Topological methods in surface dynamics.
\newblock Topology Appl. 1994;58:223--298.

\bibitem{Yomdin1987}
Yomdin Y.
\newblock Volume growth and entropy.
\newblock Israel J Math. 1987;57(3):285--300.

\bibitem{Newhouse1988}
Newhouse SE.
\newblock Entropy and volume.
\newblock Ergod Th Dynam Sys. 1988;8:283--299.

\bibitem{Newhouse1993}
Newhouse SE, Pignataro T.
\newblock On the estimation of topological entropy.
\newblock J Stat Phys. 1993;72(5-6):1331--1351.

\bibitem{Burau1936}
Burau W.
\newblock \"{U}ber {Z}opfgruppen und gleichsinnig verdrilte {V}erkettungen.
\newblock Abh Math Semin Hamburg Univ. 1936;11:171--178.

\bibitem{Fried1986}
Fried D.
\newblock Entropy and twisted cohomology.
\newblock Topology. 1986;25(4):455--470.

\bibitem{Kolev1989}
Kolev B.
\newblock Entropie topologique et repr\'{e}sentation de {B}urau.
\newblock C R Acad Sci S\'{e}r {I}. 1989;309(13):835--838.
\newblock English translation at {\tt http://arxiv.org/abs/math.DS/0304105.}

\bibitem{BandBoyland2007}
Band G, Boyland PL.
\newblock The {B}urau estimate for the entropy of a braid.
\newblock Algeb Geom Topology. 2007;7:1345--1378.

\bibitem{Bestvina1995}
Bestvina M, Handel M.
\newblock Train-Tracks for Surface Homeomorphisms.
\newblock Topology. 1995;34(1):109--140.

\bibitem{HallTrain}
Hall T. \textit{Train: {A} {C++} program for computing train tracks of surface
  homeomorphisms};.
\newblock {\tt
  http://www.liv.ac.uk/maths/PURE/MIN\_SET/CONTENT/members/T\_Hall.html}.

\bibitem{LanneauThiffeault2011}
Lanneau E, Thiffeault JL.
\newblock On the minimum dilatation of pseudo-{A}nosov homeomorphisms on
  surfaces of small genus.
\newblock Ann Inst Fourier. 2011;61(1):105--144.

\bibitem{LanneauThiffeault2011_braids}
Lanneau E, Thiffeault JL.
\newblock On the minimum dilatation of braids on the punctured disc.
\newblock Geometriae Dedicata. 2011 Jun;152(1):165--182.

\end{thebibliography}
%\printindex[subject]
%\printindex[authors]

\end{document}